# Soft Matter

rsc.li/soft-matter-journal

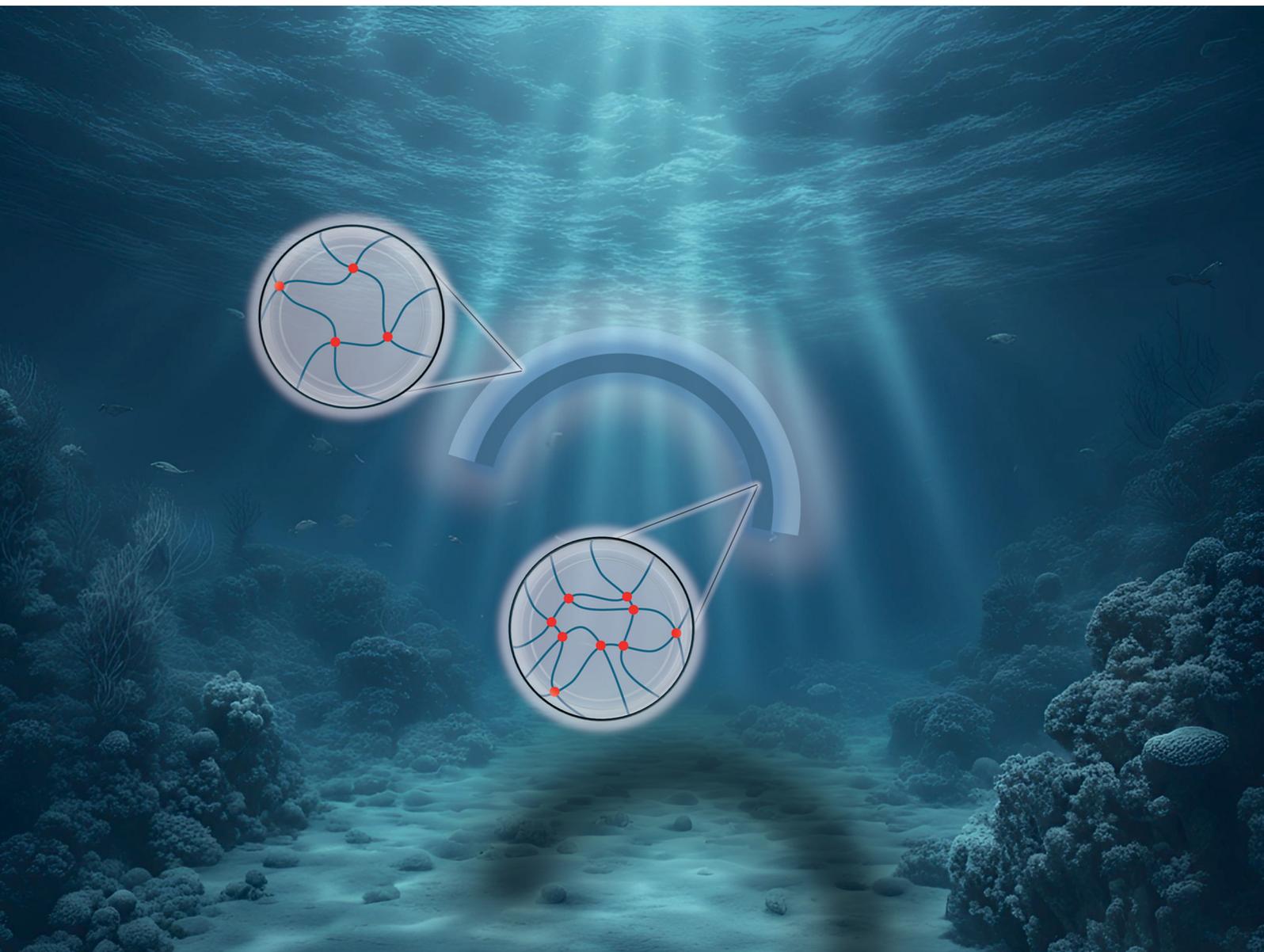

**COMMUNICATION**
Lorenzo Bonetti *et al.*
Solvent-triggered shape change in gradient-based 4D printed bilayers: case study on semi-crystalline polymer networks

ROYAL SOCIETY OF CHEMISTRY

# Solvent-triggered shape change in gradient-based 4D printed bilayers: case study on semi-crystalline polymer networks

Lorenzo Bonetti, *[a] Aron Cobianchi,[a] Daniele Natali,[b] Stefano Pandini,[c] Massimo Messori,[d] Maurizio Toselli[b] and Giulia Scalet[a]

We propose an approach to 4D print solvent-triggered, gradient-based bilayers made of semi-crystalline crosslinked polymer networks. Out-of-plane bending is obtained after immersion in the solvent, exploiting the different swelling degrees of the layers resulting from crosslinking gradients. Lastly, a beam model of the shape transformation is applied and experimentally validated.

## Introduction

Solvent-triggered polymer actuators are capable of performing mechanical work through controlled shape transformations in response to the diffusion of solvent molecules into their structure.[1,2] Given this ability, they are drawing significant attention since they find application in a broad spectrum of fields, ranging from microelectronics to soft robotics and medical devices.[3]

In this panorama, 4D printing represents a powerful and flexible technology to manufacture actuators with complex architectures and enhanced shape-morphing over time (the 4th dimension) after fabrication.[1,4] In particular, 4D printed bilayers are among the simplest and most investigated actuation systems,[5] where their shape transformation is generally an out-of-plane bending. Narrowing the focus to extrusion-based 4D printing, that offers several advantages over other printing techniques (*e.g.*, low costs, ease of use, and versatility),[6–8] bending of bilayer structures has been reported by means of different approaches, among which multi-material layering,[9] layer pre-stretching,[10] and crosslinking gradients.[11,12] In particular, the latter approach simplifies the design and fabrication steps by achieving bending using a single material whose properties vary along the thickness of the structure; this avoids the need for multi-material printers, adhesion issues, and additional setups for pre-stretch application. However, despite its advantages, this approach is still poorly investigated within the panorama of extrusion-based 4D printing and only limited to direct ink writing (DIW) of hydrogels.[11]

In this work, a versatile and simple extrusion-based 4D printing method is reported for the first time to fabricate solvent-triggered gradient-based bilayers made of semi-crystalline crosslinked polymer networks. This latter class of materials was selected as a case study, given the general interest it holds in the field of shape-memory materials.[6,13] Notably, the approach based on the crosslinking gradient here explored can be conveniently extended to other polymeric materials. Particularly, fused particle fabrication is the extrusion-based technique exploited. Alongside, a theoretical model to predict the bending behavior is applied and experimentally validated.

## Materials and methods

### Materials

PCL diol (α,ω-hydroxyl-terminated PCL, $M_n \sim$ 10 kDa), 2-isocyanatoethyl methacrylate (2-IEM, 98%), and tetrahydrofuran (THF) were purchased from Merck (Merck Life Science S.r.l., Italy), while 2-hydroxy-2-methyl-1-phenylpropanone (ADDITOL HDMAP) was purchased from Cytec. All reagents were used without further purification.

### 4D printing

Photo-crosslinkable methacrylated PCL (PCL10-MA) was obtained as previously reported,[13] then mixed with the photo-initiator (ADDITOL HDMAP, 0.5 wt%). A pneumatic bioprinter (Cellink BioX6) equipped with a thermoplastic printhead was used for 4D printing. The printing parameters, selected according to our previous work,[6] are reported in Table 1. Photo-crosslinking was achieved using the integrated UV module ($\lambda$ = 365 nm, $I$ = 0.5 mW cm$^{-2}$) of the 3D printer, irradiating each printed layer after extrusion.

[a] *Department of Civil Engineering and Architecture, University of Pavia, Via Ferrata 3, Pavia 27100, Italy. E-mail: lorenzo.bonetti@unipv.it*
[b] *Department of Industrial Chemistry "Toso Montanari", University of Bologna, Viale Risorgimento 4, Bologna 40136, Italy*
[c] *Department of Mechanical and Industrial Engineering, University of Brescia, Via Branze 38, Brescia 25133, Italy*
[d] *Department of Applied Science and Technology, Politecnico di Torino, Corso Duca degli Abruzzi 24, Torino 10129, Italy*

**Table 1** Printing parameters

| Printing parameter | Value |
| --- | --- |
| Nozzle diameter | 22 G (0.41 mm) |
| Printhead temperature | 70 °C |
| Pressure | 15 kPa |
| Speed | 10 mm s$^{-1}$ |
| Infill | 98% (concentric) |

A bilayer (30 × 5 × 0.82 mm, Fig. 2A) was fabricated controlling both the print bed temperature and UV exposure time of each layer during UV irradiation (Table 2).

Specifically, after its deposition, the bottom layer was exposed to a double UV irradiation (first: 120 s, 20 °C; second: 120 s, 60 °C) by controlling the print bed temperature. After cooling the print bed to 20 °C, the top layer was then deposited on the bottom layer and UV irradiated (120 s, 20 °C).

Single layer samples (30 × 5 × 0.41 mm) with the same UV irradiation conditions of the bottom and top layers of the bilayer were fabricated and used as control.

### Mechanical characterization

Before testing, all the 4D printed specimens were preconditioned at 60 °C and then cooled to room temperature (RT, 20 °C) to homogenize their thermal history. Quasi-static tensile tests were carried out using a Dynamic Mechanical Analyzer (DMA Q 850, TA Instruments) in the tensile configuration. Specifically, the tests were carried out at RT by applying a load ramp (1 N min$^{-1}$, 0.001 N preload) up to 18 N until break. The Young's modulus ($E$) was calculated from the slope of the stress/strain ($\sigma/\varepsilon$) curves, in the 0–0.5% strain range ($R^2 \geq 0.99$).

### Gel content and swelling

The gel content was assessed gravimetrically by first weighing the samples ($w_0$) before immersion in THF (0.5 g$_{PCL}$: 15 mL$_{THF}$ immersion ratio). Next, the samples were immersed in THF (24 h, RT), removed from the solvent, air dried, and weighted ($w_d$). The gel content ($G$ (%)) was calculated according to eqn (1):

$$G\ (\%) = \frac{w_d}{w_0} \times 100 \quad (1)$$

Similarly, swelling tests were carried out on 4D printed single layer samples through immersion in THF to get information on the dimensional variation due to swelling. The longitudinal swelling-induced strain ($\varepsilon_{swelling}$) was calculated from eqn (2):

$$\varepsilon_{swelling} = \frac{l - l_0}{l_0} \quad (2)$$

where $l_0$ and $l$ are the length of the sample before immersion and at the swelling equilibrium (24 h, RT), respectively.

### Solvent-triggered bending deformation

The bending radius ($\rho$) of the 4D printed bilayer was measured (Image J software, v. 1.53t) from images obtained with a digital camera (Canon EOS R10). The curvature ($K$) was than calculated as the inverse of the bending radius ($1/\rho$).

### Modeling of the bending deformation

To theoretically calculate the curvature of the bilayer, a modified version of the Timoshenko model was used.[14] In particular, the curvature ($K$) was here described with eqn (3), as follows:

$$K = \frac{1}{\rho} = \frac{6\left(\varepsilon_{swelling}^{bottom} - \varepsilon_{swelling}^{top}\right)(1+m)^2}{h\left(3(1+m)^2 + (1+mn)\left(m^2 + \frac{1}{mn}\right)\right)} \quad (3)$$

where $h$ is the thickness of the bilayer, $m = h_{top}/h_{bottom}$ is the thickness ratio, $n = E_{top}/E_{bottom}$ is the ratio of the Young's moduli, and $\varepsilon_{swelling}^{bottom}$ and $\varepsilon_{swelling}^{top}$ are the swelling-induced strains of the single layers. All calculations were performed with Matlab software (v. R2021b).

The relative error was calculated as follows (eqn (4)):

$$\frac{\left|K_{exp} - K_{theo}\right|}{K_{exp}} \quad (4)$$

where $K_{exp}$ and $K_{theo}$ are the mean experimental and theoretical curvatures, respectively.

### Statistical data analysis

Where possible, tests were carried out in triplicate and data expressed as mean (M) ± standard deviation (SD).

## Results and discussion

Fused particle fabrication (FPF) was the additive manufacturing technique used,[15] thus printing was achieved directly from PCL-MA pellet. Bilayer structures were successfully 4D printed in a straightforward process, exploiting a single material (PCL-MA) and varying the crosslinking degree in the different layers simply by controlling the UV light exposure time and the print bed temperature. In such a way, the variation of the

**Table 2** Layer specifications: thickness and UV crosslinking conditions

| Layer | Thickness (mm) | UV exposure time (s) | Print bed temperature (°C) |
| --- | --- | --- | --- |
| Bottom | 0.41 | 120 | 20 |
|  |  | 120 | 60 |
| Top | 0.41 | 120 | 20 |

Note: the bottom layer was obtained through double UV irradiation, varying the print bed temperature.

**Table 3** Main physical-mechanical parameters obtained for single layer structures

| Layer | $G$ (%) | $\varepsilon_{swelling}$ | $E$ (MPa) |
| --- | --- | --- | --- |
| Bottom | 92.0 ± 1.0 | 0.52 ± 0.01 | 140.0 ± 1.3 |
| Top | 84.5 ± 0.1 | 0.57 ± 0.02 | 123.4 ± 8.8 |

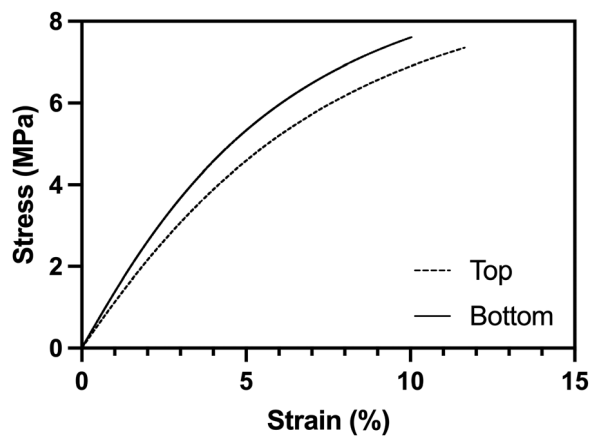

Fig. 1 Representative stress–strain curves for single layer structures. Top = dotted line, bottom = solid line.

crosslinking density along the thickness of the structure determines a gradient of the material properties and, consequently, a differential response when immersed in the solvent.

The obtained bilayer structures were thus studied for their capacity to undergo solvent-triggered shape change. For this purpose, single layer structures were first investigated from a physical-mechanical point of view. Table 3 reports the main parameters obtained from this investigation. In particular, the control of UV light exposure time and print bed temperature resulted in a control over the gel content of the two layers and, accordingly, of the swelling-induced longitudinal strain ($\varepsilon_{swelling}$) after immersion in THF.

Moreover, increased crosslinking extent resulted in increased mechanical properties (i.e., $E$) among the two layers (Table 3 and Fig. 1).

Following this investigation, the solvent-triggered behavior of the bilayer structures was explored. The 3D printed bilayer underwent out-of-plane bending (or folding) when exposed to the solvent (Fig. 2B), as a result of the crosslinking gradient generated during printing among the bottom and the top layers.

Consequently, different crosslinking extents led to different swelling degrees in the two layers, with the bottom layer swelling less than the top one. Overall, such preferential swelling led to a bending deformation towards the bottom layer (Fig. 2B). Conversely, the as-printed single layer structures immersed in the solvent did not bend at their swelling equilibrium, but simply underwent isotropic swelling-induced expansion (Table 3). This confirmed that the governing mechanism of folding of the bilayer structures was the crosslinking gradient obtained during printing.

The curvature of the bilayers structures ($K$) was then assessed as the inverse of the radius of curvature ($\rho$, Fig. 2B) via image analysis and compared with the theoretical one.

We computed the radius of curvature of the folded bilayer in the swollen state using the modified Timoshenko equation (eqn (3)). The original Timoshenko model describes the curvature of a bilayer structure composed of two materials with different coefficients of thermal expansion.[14] Here, we extended and applied such a model to a bilayer structure undergoing preferential swelling (top vs. bottom layer) due to the difference in the crosslinking degree of the two layers.[16] The quantities needed in eqn (3) were taken from Tables 2 and 3. It is worth noting that the model assumes that the two elastic moduli and layer thicknesses remain constant upon swelling, and their values are equal to those measured experimentally in the dry state. A discussion on this model assumption is available elsewhere.[17]

Interestingly, an optimal fit was obtained, with $K$ values of $0.0927 \pm 0.001$ vs. $0.0914$ mm$^{-1}$ for the measured (i.e., experimental) and theoretical cases, respectively. The relative error (eqn (4)) was equal to 0.0140. Despite the simplified assumptions, such result indicates the suitability of the adopted model in the description of the solvent-triggered shape change in the proposed bilayer for the investigated range of strains.

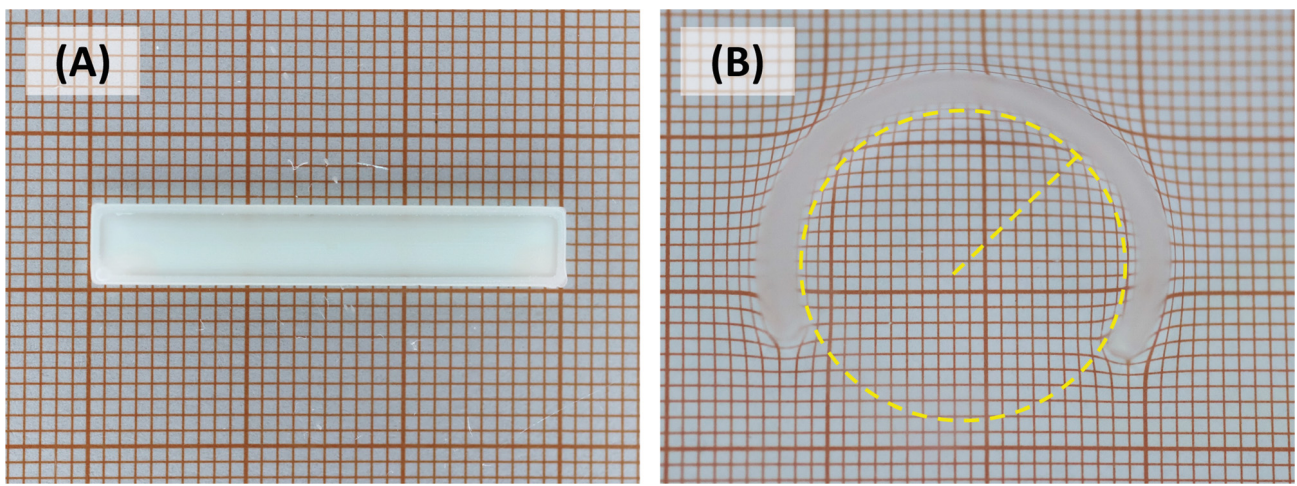

Fig. 2 Representative images of the 4D printed bilayer (A) after printing and (B) at the swelling equilibrium (24 h, RT in THF). The dotted yellow line allows to compute the radius of curvature $\rho$, used for the experimental calculation of $K$ values. Note: the swollen specimen (Fig. 2B) was rotated by 90° (along its length axis) to acquire the image.

## Conclusions

In this work, we presented, for the first time, extrusion-based 4D printed polymer bilayers consisting of a semi-crystalline polymer network capable of undergoing solvent-triggered shape change after printing. Such bilayers underwent out-of-plane bending ($K_{exp}$ = 0.0927 ± 0.001) when exposed to solvent, as a result of the crosslinking gradient generated among the printed layers during the fabrication process. In particular, crosslinking was controlled directly during the printing process leading to differences in terms of Young's modulus ($E$ = 140.0 ± 1.3 $vs.$ 123.4 ± 8.8), gel content ($G$ = 92.0 ± 1 $vs.$ = 84.5 ± 0.1), and swelling-induced strain ($\varepsilon_{swelling}$ = 0.52 ± 0.01 $vs.$ 0.57 ± 0.02) between the bottom and top layers of the 4D printed bilayers. Also, we proposed a simple model to predict the solvent-triggered bending of the 4D printed bilayer exploiting a modified version of the Timoshenko equation and validating it with experimental data. Interestingly, we obtained a low relative error (0.0140) indicating that an optimal fit was obtained.

Several different industrial applications can be envisioned for such systems, ranging from actuators responsive to organic solvents to sensors capable to respond/detect toxic or explosive chemicals ($e.g.$, liquids, vapors).[18]

Notably, this work can be easily extended to other bilayers undergoing solvent-triggered shape change, thus laying the foundations for model-assisted optimal design of different bilayers ($e.g.$, different material(s), geometrical, and physical-mechanical features) and/or of even more complex layered designs.

## Author contributions

Lorenzo Bonetti: conceptualization, data curation, formal analysis, investigation, methodology, visualization, writing – original draft, writing – review & editing. Aron Cobianchi: investigation, writing – original draft. Daniele Natali: conceptualization, writing – review & editing. Stefano Pandini: conceptualization, writing – review & editing. Massimo Messori: conceptualization, writing – review & editing. Maurizio Toselli: conceptualization, writing – review & editing. Giulia Scalet: conceptualization, data curation, formal analysis, funding acquisition, investigation, methodology, project administration, supervision, visualization, writing – original draft, writing – review & editing.

## Data availability

Data generated in this study are deposited in the Zenodo database at **https://doi.org/10.5281/zenodo.11110714**.

## Conflicts of interest

There are no conflicts to declare.

## Acknowledgements

This work was funded by the European Union ERC CoDe4Bio Grant ID 101039467. Views and opinions expressed are however those of the author(s) only and do not necessarily reflect those of the European Union or the European Research Council. Neither the European Union nor the granting authority can be held responsible for them.

## Notes and references

1 S. Parimita, A. Kumar, H. Krishnaswamy and P. Ghosh, $J.$ $Manuf.$ $Process.$, 2023, **85**, 875–884.
2 J.-W. Su, X. Tao, H. Deng, C. Zhang, S. Jiang, Y. Lin and J. Lin, $Soft$ $Matter$, 2018, **14**, 765–772.
3 P. Martins, D. M. Correia, V. Correia and S. Lanceros-Mendez, $Phys.$ $Chem.$ $Chem.$ $Phys.$, 2020, **22**, 15163–15182.
4 X. Dong, X. Luo, H. Zhao, C. Qiao, J. Li, J. Yi, L. Yang, F. J. Oropeza, T. S. Hu, Q. Xu and H. Zeng, $Soft$ $Matter$, 2022, **18**, 7699–7734.
5 P. Mainik, L. Hsu, C. W. Zimmer, D. Fauser, H. Steeb and E. Blasco, $Adv.$ $Mater.$ $Technol.$, 2023, **8**, 2300727.
6 L. Bonetti, D. Natali, S. Pandini, M. Messori, M. Toselli and G. Scalet, $Mater.$ $Des.$, 2024, **238**, 112725.
7 D. Rahmatabadi, K. Soltanmohammadi, M. Aberoumand, E. Soleyman, I. Ghasemi, M. Baniassadi, K. Abrinia, M. Bodaghi and M. Baghani, $Phys.$ $Scr.$, 2024, **99**, 025013.
8 D. Rahmatabadi, M. Aberoumand, K. Soltanmohammadi, E. Soleyman, I. Ghasemi, M. Baniassadi, K. Abrinia, M. Bodaghi and M. Baghani, $Adv.$ $Eng.$ $Mater.$, 2023, **25**, 2201309.
9 P. Cao, J. Yang, J. Gong, L. Tao, T. Wang, J. Ju, Y. Zhou, Q. Wang and Y. Zhang, $J.$ $Appl.$ $Polym.$ $Sci.$, 2023, **140**, e53241.
10 G. Qu, J. Huang, Z. Li, Y. Jiang, Y. Liu, K. Chen, Z. Xu, Y. Zhao, G. Gu, X. Wu and J. Ren, $Mater.$ $Today$ $Bio$, 2022, **16**, 100363.
11 P. Cao, L. Tao, J. Gong, T. Wang, Q. Wang, J. Ju and Y. Zhang, $ACS$ $Appl.$ $Polym.$ $Mater.$, 2021, **3**, 6167–6175.
12 J. Odent, S. Vanderstappen, A. Toncheva, E. Pichon, T. J. Wallin, K. Wang, R. F. Shepherd, P. Dubois and J.-M. Raquez, $J.$ $Mater.$ $Chem.$ $A$, 2019, **7**, 15395–15403.
13 N. Inverardi, M. Toselli, G. Scalet, M. Messori, F. Auricchio and S. Pandini, $Macromolecules$, 2022, **55**, 8533–8547.
14 S. Timoshenko, $J.$ $Opt.$ $Soc.$ $Am.$, 1925, **11**, 233.
15 A. Bayati, D. Rahmatabadi, I. Ghasemi, M. Khodaei, M. Baniassadi, K. Abrinia and M. Baghani, $Mater.$ $Lett.$, 2024, **361**, 136075.
16 V. Stroganov, M. Al-Hussein, J.-U. Sommer, A. Janke, S. Zakharchenko and L. Ionov, $Nano$ $Lett.$, 2015, **15**, 1786–1790.
17 Y. Wu, X. Hao, R. Xiao, J. Lin, Z. L. Wu, J. Yin and J. Qian, $Acta$ $Mech.$ $Solida$ $Sin.$, 2019, **32**, 652–662.
18 H. Lin, S. Zhang, Y. Xiao, C. Zhang, J. Zhu, J. W. C. Dunlop and J. Yuan, $Macromol.$ $Rapid$ $Commun.$, 2019, **40**, 1800896.